\documentclass[aps,pra,twocolumn,floats,showpacs,superscriptaddress,10pt]
{revtex4-1}

\usepackage{bm}
\usepackage{graphicx}% Include figure files
\usepackage{amssymb}
\usepackage{amsfonts}
\usepackage{amsmath}
\usepackage{epstopdf}
\usepackage{color}

\def \IFPAN{Institute of Physics, Polish Academy of Sciences, al. 
Lotnik\'{o}w 32/46, 02-668 Warsaw, Poland}
\def \KWU{Institute of Physics, Kazimierz Wielki University, Powstanc\'{o}w Wielkopolskich 2, 85-064 Bydgoszcz, Poland}

\begin{document}

\title{Transition metal ions in ZnO: effects of intrashell Coulomb repulsion on electronic properties}

\author{A. Ciechan}\email{ciechan@ifpan.edu.pl}\affiliation{\IFPAN}
\author{P. Bogus{\l}awski}\email{bogus@ifpan.edu.pl}\affiliation{\IFPAN,\KWU}
 
\date{\today}
%&&&&&&&&&&&&&&&&&&&&&&&&&&&&&

\begin{abstract}
Electronic structure of the transition metal (TM) dopants in ZnO is calculated by first principles approach. Analysis of the results is focused on the properties determined by the intrashell Coulomb coupling. The role of both direct and exchange interaction channel is analyzed. The coupling is manifested in the strong charge state dependence of the TM gap levels, which leads to the metastability of photoexcited Mn, and determines the accessible equilibrium charge states of TM ions. 
The varying magnitude of the exchange coupling is reflected in the dependence of the spin splitting energy on the chemical identity across the 3$d$ series, as well as the charge state dependence of spin-up spin-down exchange splitting.  
\end{abstract}


\maketitle

\section{\label{sec1}Introduction}
Current interest in ZnO doped with various dopants stems from the recognized potential of ZnO in various applications. 
In particular, the usage of transition metal (TM) impurities was considered in various contexts, including both tunable magnetic properties and improved optical/transport properties~\cite{Ando1, Ando2, Kittilstved, PacuskiCo, PacuskiMn}. A representative example is application of ZnO doped with Mn for photocatalysis~\cite{Maeda}. A trivial statement is that controlled applications require a good understanding of electronic properties of the given dopant. 

The present study is devoted to selected theoretical aspects of the physics of TMs dopants in ZnO. Our attention is focused on the role of the intrashell Coulomb coupling. To highlight characteristic features, we show the results obtained for the TM series from Ti to Cu, and discuss the trends displayed by their electronic structure, i.e., by energy levels, spin states, and accessible charge states. In a many body system, electron-electron coupling is realized through the direct as well as the exchange channel, the latter being a result of the antisymmetrization of the wave functions, and leading to the Hund's rule. Both channels determine the electronic structure of TM ions in ZnO. 

Majority of TM ions induce states in the band gap of the host ZnO crystal. The "deep" character of those states is reflected in the structure of the wave functions, which are mainly built up from the localized atomic-like $d$ orbitals of the defect, while the contribution of the orbitals of its neighbors is smaller. Spacial localization in turn implies that the Coulomb repulsion between electrons occupying impurity levels is pronounced, and affects level energies by about 1~eV. 
(This is in contrast to the shallow and extended states, where the Coulomb energies are of the order of 0.01~eV.) In most cases, a TM impurity can assume several charge states depending on the Fermi energy, which differ by the number of electrons that occupy impurity levels, and thus by the varying Coulomb intrashell coupling. In particular, with the increasing (decreasing) occupation of the defect gap level, its energy increases (decreases) as a result of the increased (decreased) Coulomb repulsion. 
This effect determines the possible charges states of the TM ion in ZnO. 

A second consequence of the intrashell Coulomb coupling discussed below is the charge state dependence of the exchange coupling, revealed by the magnitude of the spin splitting of the TM levels.  
Finally, the third consequence of the strong localization of the wave function of the gap states is that the change of the charge state induces large variations of the distance between the TM ion and its nearest neighbors, which are of the order of a few percent. Such atomic relaxations affect the gap levels by about 1~eV. 

Experimental investigations of TM in ZnO include optical and transport studies. Optical transitions observed in both absorption and emission can be either intracenter, i.e. between two levels of the TM ion, in which case the charge state of TM does not change, or the ionization transition, which excites an electron from the center to conduction band minimum (CBM), or from the valence band maximum (VBM) to the TM level. In the case of intracenter transition, the bond length between TM and its neighbors remain practically the same, and the transition energy is well approximated by the difference of the one electron eigenenergies of the gap states. On the other hand, ionization transitions imply pronounced atomic relaxations, and non-negligible Frank-Condon effect.

%%%%%%%%%%%%%%%%%%%%%%%%%%%%%%%%%%%%%%%%%%%%%
\section{\label{sec2}Method of Calculations}
The calculations are done by employing the generalized gradient approximation (GGA) to the density functional theory~\cite{Hohenberg,KohnSham,PBE}, supplemented by the $+U$ corrections~\cite{Anisimov1991, Anisimov1993, Cococcioni} meant to improve the agreement with experiment. We use the pseudopotential method implemented in the QUANTUM ESPRESSO code~\cite{QE} with the valence atomic configuration $3d^{10} 4s^2$ for Zn, $2s^2p^4$ for O and $3s^2p^6 4s^2p^0 3d^n$ or $4s^2p^0 3d^n$ for TM ions with $n$ electrons on $d$(TM) shell. The plane-waves kinetic energy cutoffs of 30~Ry for wave unctions and 180~Ry for charge density are employed. The electronic structure of the wurtzite ZnO is examined with a $8\times 8\times$ 8 $k-$point grid. Analysis of a single TM impurity substituting for the Zn lattice is performed using $3\times 3\times 2$ supercells with 72 atoms, while $k-$space summations are performed with a $3\times 3\times$ 3 $k-$point grid. 

Theoretical description of the electronic structure of TM ions in semiconductors must begin with an accurate band structure of the host. In fact, underestimation of the band gap, typical for the local density approximation (LDA) or GGA will result in erroneous properties of dopants. 
For example, an underestimated band gap leads to a distorted electronic structure of TM when the TM state is predicted to be resonant with the continuum of the conduction band instead of being the gap state. 
As a consequence, wrong optical transition energies, metallic rather than insulating conductivity, and a non-correct type of magnetic coupling between TM atoms are obtained~\cite{Raebiger, Lany_PRB77}. 
Here, a correct band gap is obtained by employing the $+U$ corrections to the GGA calculations~\cite{Mn, Fe}. Application of $+U$ correction only to $d$(Zn)~\cite{fang, ye-theo} partially improves the situation.
We find that applying the $U$ correction also to $p$(O) orbitals, $U$(O)=6.25~eV, in addition to $U$(Zn)=12.5~eV, gives not only the experimental $E_{gap}$ of 3.3~eV~\cite{Dong, Izaki:APL1996, Srikant:JAP1998} but also the energy of the $d$(Zn) band, centered about 8~eV below the VBM~\cite{Lim}. Our $U$ parameters for ZnO are similar to the values reported in other works~\cite{Ma,Calzolari,marco}. 
Interestingly, the usage of hybrid functionals render a correct $E_{gap}$, but the agreement with experimental data for TM ions is far from satisfactory~\cite{Zakrzewski}. 

From our previous results it follows that the $U$ corrections for the $d$(TM) orbitals are smaller than that used for $d$(Zn) in pure ZnO. While the inclusion of $U$(TM) improves the agreement with experiment, it turns out that the fitted $U$ values are relatively small, and so are the energy changes. Typically, agreement with experiment is obtained for $U$(TM) of about  $2-3$~eV \cite{Zakrzewski, Mn, Fe}. A non-vanishing $U$(TM) implies occupation dependent corrections to the TM levels~\cite{Mn, Fe}. 
Since a comparison with the existing experimental data is beyond the scope of this work, we keep the discussion more transparent and highlight the impact of the Coulomb coupling by neglecting $U$(TM). 
The actual values of the $U$ term for TM ions in ZnO were optimized by fitting the energies of ionization and intracenter optical transitions to experimental data~\cite{Mn,Fe}.

Transition level between various charge states of a defect is defined as the Fermi energy at which formation energies of these states are equal. Our calculations are performed along the scheme proposed in Ref.~\cite{Lany_PRB78} including the image charge corrections and potential alignment for charged defects~\cite{Lany_PRB78,Lany_ModelSim17, Freysoldt}. 

The ionization energies are calculated as the total energy difference between the final and the initial states of the system. The thermal ionization energy can be obtained directly from the position of the (+/0) level relative to CBM. The calculation of absorption-recombination transitions for Mn ion are performed with fixed occupation matrices at the $\Gamma$ point~\cite{Mn}. In this case, the finite size corrections to the total energy are not necessary.

%%%%%%%%%%%%%%%%%%%%%%%%%%%%%%%%%%%%%%%%%%%%%
\section{\label{sec3}Electronic Structure od TM ions in ZnO}

\begin{figure}[!]
\begin{center}
\includegraphics[width=8.3cm]{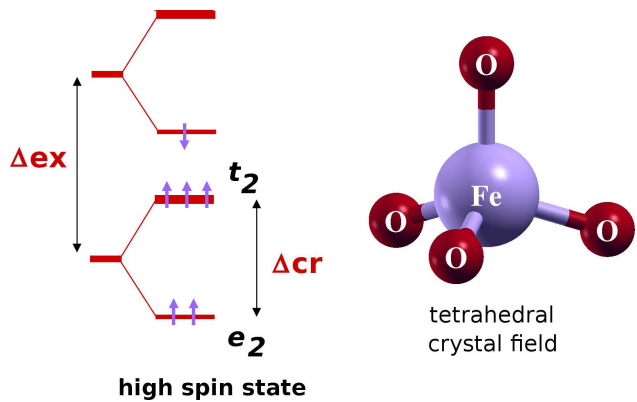}
\end{center}
\caption{\label{fig1} 
Left panel: Schematics of the TM levels in ZnO, Fe$^{2+}$ is taken as an example. Electron spins are shown by arrows, doublet (triplet) levels are denoted by thin (thick) lines. Right panel: the tetrahedrally arranged oxygen neighbors of Fe; in the wurtzite structure the three planar bonds are not equivalent with the vertical one. 
}
\end{figure}

\begin{figure}[t!]
\begin{center}
\includegraphics[width=8.3cm]{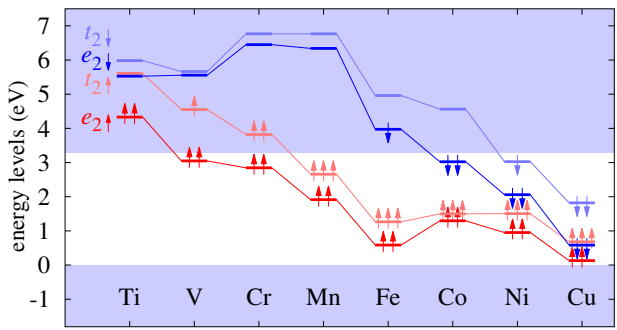}
\end{center}
\caption{\label{fig2} 
Calculated $d$(TM) levels assuming $U$(TM)=0. Small splitting of $t_2$ triplets is neglected. Arrows denote spins.
}
\end{figure}

We begin by recalling that the electronic structure of a TM in a crystal follows from two effects. The first one is the exchange splitting of the $d$-electrons into spin up and spin down states following the Hund's rule. The second effect is the splitting of the $d$(TM) orbital quintet by the crystal field into a doublet and a triplet distant by about 1~eV. 
In the wurtzite ZnO, the latter is weakly split into a singlet and a doublet, typically by about 0.1~eV. This splitting is neglected in Figs~\ref{fig1} and \ref{fig2}. The level structure is schematically shown in Fig.~\ref{fig1}.  

Both exchange and crystal field splittings are clearly visible in Fig.~\ref{fig2}, which presents the calculated one electron energies of neutral TM$^{2+}$ ranging from Ti to Cu. 
The relative magnitudes of splittings depend on the chemical identity of the ion, and are addressed below. 
In all cases, the exchange splitting exceeds the crystal field splitting, and all the TM ions are in the high spin state. 
Next, as it follows from the figure, with the increasing nuclear charge of TM, the levels decrease in energy ~\cite{note1}. Previous calculations~\cite{Raebiger} gave results similar to ours.

%%%%%%%%%%%
\subsection{\label{sce3a} The dependence of the impurity levels on the charge state}
\begin{figure}[t!]
\begin{center}
\includegraphics[width=8.3cm]{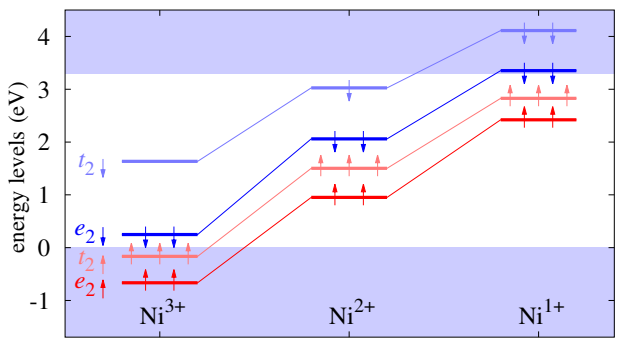}
\end{center}
\caption{\label{fig3} 
Charge state dependence of the Ni levels. Arrows denote spins. In the case of Ni$^{3+}$, the spin up Ni states are strongly hybridized with the valence bands, and their energies are shown only schematically. 
}
\end{figure}

The effect is exemplified in Fig.~\ref{fig3} for Ni. One can see that adding or subtracting one electron from $t_{2\downarrow}$ changes its energy by as much as $\sim$1.5~eV. This demonstrates the magnitude of the intrashell Coulomb coupling, and makes the 1+ charge state of Ni unstable. Indeed, the capture of an electron on the $e_{2\downarrow}$ level increases its energy above the CBM, i.e., $e_{2\downarrow}$ becomes a resonance degenerate with the conduction band continuum. In such a case, autoionization of Ni$^{1+}$ is expected to occur. 
On the other hand, removal of an electron 
from $e_{2\downarrow}$ of Ni$^{2+}$ results in Ni$^{3+}$ charge state which is stable. However, one can expect that the removal of the second electron further lowers 
the Ni gap states, and Ni$^{4+}$ charge state cannot be reached because $e_{2\downarrow}$ would merge with the valence band. 
This shows that the Coulomb coupling largely limits the accessible charge states of a TM impurity. Similar results are obtained for other dopants.

\begin{figure}[t!]
\begin{center}
\includegraphics[width=8.3cm]{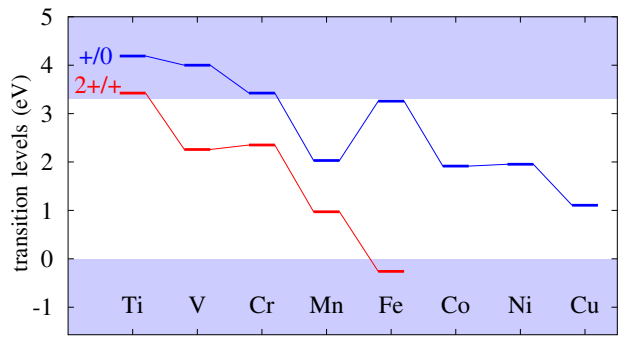}
\end{center}
\caption{\label{fig4} 
Transition levels of TM in ZnO. 
}
\end{figure}

The presence of several one electron levels in the band gap
(four levels in the case of Ni$^{2+}$, see Fig.~\ref{fig2}) suggests that an ion can assume many charge states. 
Actually, however, this is not the case, since the strong Coulomb coupling limits the number of equilibrium charge states, as explained for Ni. 
The possible charge states of TM ions are given by transition levels, which are shown in Fig.~\ref{fig4}. 
In all cases except Ti and Mn, only two charge states can be assumed. 
One electron energies of Fig.~\ref{fig2} can be investigated in optical experiments, which allow to observe intracenter transitions between the gap levels, while the optically or thermally driven ionization transitions are given by the transition levels shown in Fig.~\ref{fig4}.

%%%%%%%%%%%
\subsection{\label{sce3b} Exchange splitting}
\begin{figure}[t!]
\begin{center}
\includegraphics[width=8.3cm]{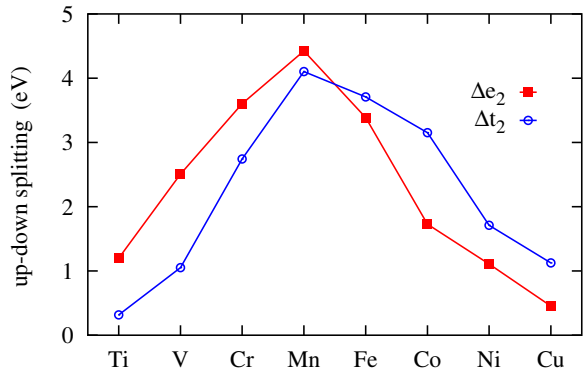}
\end{center}
\caption{\label{fig6} 
The calculated exchange spin-up spin-down splitting of $e_2$ doublets (squares) and $t_2$ triplets (dots) for TMs in ZnO. Note that in both cases the maximal value is achieved for Mn. 
}
\end{figure}
A second manifestation of the intrashell coupling is provided by the exchange induced spin-up spin-down splittings of the $d$(TM) levels, denoted as $\Delta ex$ in Fig.~\ref{fig1}. 
As pointed out above, the strong charge state dependence of one electron energies follows from the direct Coulomb interaction. From Fig.~\ref{fig2} it follows that also the exchange coupling significantly affects the energy levels, or more precisely, the exchange splitting of the TM states. 

The calculated values of spin splitting of both $e_2$ and $t_2$ states, presented in Fig.~\ref{fig6}, exhibit a strong dependence on the TM ion identity. Indeed, they increase almost linearly from about 1~eV for Ti, to $\sim$5~eV for Mn, and then decrease to about 1~eV for Cu. The splittings of the $e_2$ and $t_2$ states are comparable. 
A qualitative description of those data is obtained by assuming a simple electron-electron coupling of the Heisenberg form, $H=- 2J\sum_{i,j>i}\vec{s_i}\cdot\vec{s_j}$, where the summations goes over the occupied one electron states with the spin $\vec{s_i}$. The five fold increase of the splitting from Ti to Mn is accounted for in this model when $J$ is about 1~eV. 

The same argument explains the charge state dependence of the spin splitting. For example, we find that the splitting is reduced by about 1~eV, from $\Delta e_2=4.4$~eV   ($\Delta t_2=4.1$~eV) for Mn$^{2+}$ to $\Delta e_2=3.8$~eV   ($\Delta t_2=3.1$~eV)  for Mn$^{3+}$, which corresponds to an approximately linear scaling with the number of electrons. Similar results are obtained for Cr. In this case, the change of the charge state from Cr$^{2+}$ to Cr$^{3+}$ induces the change of the spin splitting from $\Delta e_2=3.6$~eV   ($\Delta t_2=2.7$~eV)  to $\Delta e_2=3.3$~eV ($\Delta t_2=2.1$~eV). 

%%%%%%%%%%%
\subsection{\label{sce3c} Examples}
{\bf Titanium}. 
If the last nominally occupied level of both neutral and ionized TM ion is degenerate with the conduction band, a spontaneous autoionization takes place. Such a situation occurs for Ti, for which the $e_{2\uparrow}$ level is above the bottom of the conduction band not only for Ti$^{2+}$ but also for Ti$^{3+}$ and Ti$^{4+}$, as reflected by the transition level 2+/+ being above CBM, see Fig.~\ref{fig4}. Therefore, after autoionization of two electrons to CBM, Ti is in its only 
possible charge state 4+. 
This is in agreement with experiment, which shows a decrease of resistivity upon Ti doping~\cite{Yang-Ming_2004, Singh_2008, Bergum, Shao_JAP2015}.

\begin{figure}[t!]
\begin{center}
\includegraphics[width=8.3cm]{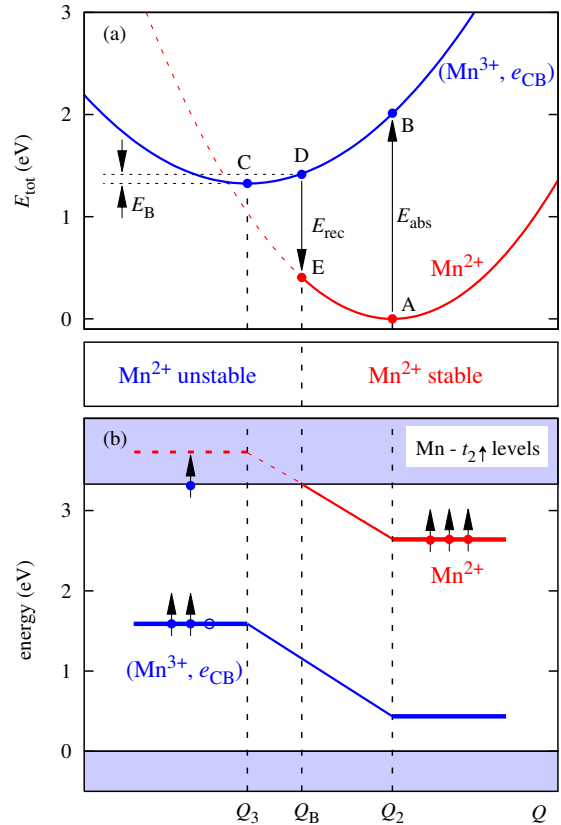}
\end{center}
\caption{\label{fig5} 
(a) Total energy of Mn$^{2+}$ and (Mn$^{3+}$, $e_{CB}$) as a function of configuration coordinate $Q$. $Q_2$ and $Q_3$ are equilibrium atomic configurations of Mn$^{2+}$ and (Mn$^{3+}$ , $e_{CB}$) charge states, respectively, and $Q_B$ is the configuration coordinate of the barrier. (b) Single particle energy of the $t_{2\uparrow}$ level for both Mn$^{2+}$ (red symbols) and (Mn$^{3+}$, $e_{CB}$) (blue symbols); note the strong dependence of the $t_{2\uparrow}$ energy on the charge state. $U$(Mn) = 0 is assumed. Arrows denote electron spins. Results of Ref.~\cite{Mn} (with permission).
}
\end{figure}

{\bf Manganese}. 
An interesting example of the role of the intracenter Coulomb repulsion is provided by Mn in ZnO, which exhibits a metastable behavior~\cite{Mn}. According to low temperature experiments, after illumination of ZnO:Mn, the photoionized Mn$^{3+}$ is in a metastable atomic configuration, in which a direct recombination of the photoelectron from CBM to Mn is not possible. The metastability follows from the large intrashell Coulomb repulsion. The ionization-recombination cycle is presented in Fig.~\ref{fig5}, where total system energy for both Mn$^{2+}$ and Mn$^{3+}$ is shown in the panel (a), and the one electron state $t_{2\uparrow}$ in panel (b). Note that the panel (b) shows the same $t_{2\uparrow}$ level for two charge states and for two atomic configurations. $Q_2$ is the ground state atomic configuration of Mn$^{2+}$. For a detailed comparison with experimental data see~\cite{Mn}.

After photoionization of Mn$^{2+}$, the ionized Mn$^{3+}$ relaxes to the equilibrium configuration $Q_3$, in which the Mn-O bond lengths decrease by about 7 per cent. In spite of the fact that the $t_{2\uparrow}$ level is below the CBM, the electron cannot recombine because in this configuration the energy of $t_{2\uparrow}$(Mn$^{2+}$) occupied by three electrons is above CBM. The energy of $t_{2\uparrow}$(Mn$^{2+}$) is below CBM when the barrier configuration $Q_B$ is reached. After recombination, Mn$^{2+}$ relaxes to $Q_2$. 

Figure 6b shows another typical consequence of the localization of $d$(TM) orbitals, which is a large, a few per cent, change of the TM-O bonds with the nearest oxygen anions induced by the change of the charge state. This in turn shifts TM gap levels.  
For example, in the case of Mn, photoionization reduces the intrashell Coulomb interaction, i.e., it decreases the $t_{2\uparrow}$ level by about 2 eV when the atomic configuration is fixed at $Q_2$. After ionization, 
the reduction of the Mn-O bonds causes the increase of $t_{2\uparrow}$ level by about 1 eV because of antibonding character of the $t_{2\uparrow}$ wavefunction. 
The two effects are comparable, but they act in the opposite way.

{\bf Iron}. 
A problem of similar nature is present for Fe$^{2+}$. 
As it was discussed in Ref. ~\cite{Fe}, there are electronic configurations, which can not be considered within DFT, because the ground state of the system is predicted for an unphysical partial occupation of two states. 
This is the case of Fe$^{2+}$, for which the $e_{2\downarrow}$ level should be occupied with one electron. 
When $U<2$ eV is assumed, the energy of $e_{2\downarrow}$ is degenerate with the continuum of the conduction band, and autoionization is expected. 
(Note that a good agreement with the experimental data is obtained for $U=4$ eV ~\cite{Fe}.) 
However, after ionization to CBM, the energy of $e_{2\downarrow}$ strongly decreases, and the empty $e_{2\downarrow}$ is below CBM. 
Numerically, the equilibrium electronic configuration is found when the electron is split between $e_{2\downarrow}$ and CBM, but such a configuration is unphysical. 

The same unphysical situation can occur for other TM do\-pants, which nominally are resonant donors with levels degenerate with the conduction band and close to the CBM. In contrast, when the $d$(TM) level is sufficiently high, i.e., in the case when the level is a resonance even after the autoionization, the properties of the center and the allowed charge states are correctly described by theory. This is the case of Ti. 

%%%%%%%%%%%%%%%%%%%%%%%%%%%%%%%%%%%%%%%%%%%%%
\section{\label{sec4}Conclusions}

Analysis of results of ab initio calculations of the electronic structure of TM ions in ZnO was conducted in order to reveal the role of the intrashell Coulomb coupling. Both one electron energies and thermodynamic transition levels were obtained. Discussion was performed for TMs ranging from Ti to Cu.  
Typically, TM ions in ZnO induce a rich spectrum of several $d$-derived one electron levels in the band gap. In spite of this, in most cases the number of possible charge states is limited to two. 
This is a result of the strong intrashell coupling, due to which gap levels shift by about 1-2~eV with the varying charge state of the ion. 
The strong intrashell Coulomb coupling explains in particular the observed metastability of optically excited Mn$^{3+}$.  
Similarly, because of the pronounced intrashell exchange coupling, 
the magnitude of the spin splitting of the $d$ electrons depends not only on the ion chemical identity, but also on its charge state.
The charge state dependence of atomic configurations of ions is also pointed out. 
Details provided for Ti, Mn and Fe were aimed to illustrate typical characteristics of TM ions in ZnO.

%%%%%%%%%%%%%%%%%%%%%%%%%%%%%%%%%%%%%%%%%%%%%
\section*{Acknowledgments}
This work is partially supported by the National Science Center of Poland based on Decisions No. 2016/21/D/ST3/03385. Calculations are performed on ICM supercomputers of Warsaw University (Grants No. G46-13 and No. G16-11).

%%%%%%%%%%%%%%%%%%%%%%%%%%%%%%%%%%%%%%%%%%%%%
\section*{References}
%\bibliographystyle{elsarticle-num}
%\bibliography{ZnObiblio_gdansk}

\begin{thebibliography}{10}
\bibitem{note1}
This dependence is violated for the pair Fe-Co. Indeed, both $t_{2\uparrow}$
  and $e_{2\uparrow}$ of Fe are below those of Co. The reason is the following.
  In the case of Fe, the sixth $d$ electron does not occupy the
  $e_{2\downarrow}$ doublet which is degenerate with the conduction band
  continuum, but autoionizes to the CBM. Consequently, its wave function is
  delocalized, which reduces the intrashell coupling, and lowers the energies
  of $t_{2\uparrow}$ and $e_{2\uparrow}$. In the case of Co, sixth and seventh
  electron are located on the localized $e_{2\downarrow}$ gap state, the
  repulsion is strong, and consequently Co levels are higher than those of Fe.
\end{thebibliography}

\end{document}